\documentclass[a4paper,11pt]{article}
\pdfoutput=1
\usepackage[english]{babel}

\usepackage[a4paper,tmargin=3truecm,bmargin=3truecm,rmargin=3truecm,
lmargin=3truecm,twoside,verbose=true]{geometry}

\usepackage{cancel,graphicx}
\usepackage{hyperref}

\usepackage{amsmath,amssymb}

\numberwithin{equation}{section}
\allowdisplaybreaks[1]

\usepackage[amsmath, hyperref, thmmarks]{ntheorem}

\usepackage{cancel,graphicx}
\usepackage{hyperref}
\numberwithin{equation}{section}

\allowdisplaybreaks[1]

\newcommand{\be}{\begin{equation}}
\newcommand{\ee}{\end{equation}}

\newcommand\bbone{{ \mathbb{I}}}

\newcommand{\deq}{: \hspace{-0.5pt} =}
\newcommand{\dreq}{= \hspace{-0.5pt} :}

\theoremsymbol{}
\theorembodyfont{\slshape}
\theoremheaderfont{\normalfont\bfseries}
\theoremseparator{}

\theorembodyfont{\upshape}
\theoremsymbol{\ensuremath{\blacklozenge}}

\theoremstyle{nonumberplain}
\theoremheaderfont{\scshape}
\theorembodyfont{\normalfont}
\theoremsymbol{\ensuremath{\blacksquare}}

\qedsymbol{\ensuremath{_\blacksquare}}
\theoremclass{LaTeX}
\renewenvironment{thebibliography}[1]
         {\section*{References}\frenchspacing\small
          \begin{list}{[\arabic{enumi}]}
         {\usecounter{enumi}\parsep=2pt\topsep 0pt
         \settowidth{\labelwidth}{[#1]}
         \leftmargin=\labelwidth\advance\leftmargin\labelsep
         \rightmargin=0pt\itemsep=1pt\sloppy}}{\end{list}}
\title{Quantum spaces, central extensions of Lie groups \\
and related quantum field theories}

\author{Timoth\'e Poulain, Jean-Christophe Wallet}

\begin{document}
\date{}
\maketitle
\begin{center}

\textit{Laboratoire de Physique Th\'eorique, B\^at.\ 210\\
CNRS and Universit\'e Paris-Sud 11,  91405 Orsay Cedex, France}\\
 \href{mailto:timothe.poulain@th.u-psud.fr}{\texttt{timothe.poulain@th.u-psud.fr}}, \href{mailto:jean-christophe.wallet@th.u-psud.fr}{\texttt{jean-christophe.wallet@th.u-psud.fr}}\\[1ex]%
\end{center}


\maketitle

\begin{abstract} 
Quantum spaces with $\frak{su}(2)$ noncommutativity can be modelled by using a family of $SO(3)$-equivariant differential $^*$-representations. The quantization maps are determined from the combination of the Wigner theorem for $SU(2)$ with the polar decomposition of the quantized plane waves. A tracial star-product, equivalent to the Kontsevich product for the Poisson manifold dual to $\mathfrak{su}(2)$ is obtained from a subfamily of differential $^*$-representations. Noncommutative (scalar) field theories free from UV/IR mixing and whose commutative limit coincides with the usual $\phi^4$ theory on $\mathbb{R}^3$ are presented. A generalization of the construction to semi-simple possibly non simply connected Lie groups based on their central extensions by suitable abelian Lie groups is discussed. 
\end{abstract}
\vskip 10 true cm
{\it{Based on a talk presented by one of us (T. Poulain) at the XXVth International Conference on Integrable Systems and Quantum symmetries (ISQS-25), Prague, June 6-10 2017.}}\\
\vfill\eject

\section{Introduction}

In the recent years, deformations of $\mathbb{R}^3$ for which the algebra of coordinates forms a $\mathfrak{su}(2)$ Lie algebra have received some interest. This was in particular related either to developments in $3$-d gravity, in particular viewing $\mathbb{R}^3$ as the dual algebra of the relativity group \cite{ritals1, ritals2}, or to constructions of analytic formulas for the star-products \cite{fedele-vitale} defining these deformations as well as investigations of their structural properties \cite{lagraa,selene,KV-15,q-grav} (see also \cite{galuccio,fedele}). Field theories built on these noncommutative (quantum) spaces have been shown to have a perturbative quantum behaviour different from the one of field theories built on Moyal spaces, at least regarding renormalizability as well as UV/IR mixing \cite{vitwal,vit-kust,wal-16}. For earlier works on noncommutative field theories (NCFT) on Moyal spaces, see e.g \cite{Grosse:2003aj-pc,Blaschke:2009c} and references therein.\\

The purpose of this paper is to select salient features developed in our recent works \cite{jpw-16,tp-jcw17} on quantum spaces with $\frak{su}(2)$ noncommutativity, hereafter denoted generically by $\mathbb{R}^3_\theta$ (see below). It appears that these latter can be modelled conveniently by exploiting a family of $SO(3)$-equivariant differential $^*$-representations as we will show in a while. It turns out that the use of differential representations \cite{polydiff-alg,VK,KV-15} may prove useful in the construction of star-products whenever the noncommutativity is of a Lie algebra type such as the case considered here. Consistency of the construction definitely requires that one works with $^*$-representations. Note that a similar construction can also be applied to the kappa-Minkowski spaces which are related to (the universal enveloping algebra of) a solvable Lie algebra. For constructions of star-products within kappa-Minkowski spaces, see e.g \cite{cosmogol4}.\\

The characterization of the related quantization maps defining the quantum spaces $\mathbb{R}^3_\theta$ can be achieved from a natural combination of the polar decomposition of the quantized plane waves with the Wigner theorem for $SU(2)$. Recall that the quantized plane waves are defined by the action of the quantization map on the usual plane waves. From this follows the characterization of the star-products. The use of star-product formulation of NCFT is convenient for fast construction of functional actions. However, it may lead to difficulties whenever the star-product is represented by a complicated formula and/or is not closed for a trace functional. In this respect, we then construct a tracial star-product with respect to the usual Lebesgue measure on $\mathbb{R}^3$, equivalent to the Kontsevich product \cite{deform} for the Poisson manifold dual to $\mathfrak{su}(2)$, thanks to a suitable use of the Harish-Chandra map \cite{harish, duflo}. We then present noncommutative (scalar) field theories which are free from UV/IR mixing and whose commutative limit is the usual $\phi^4$ theory on $\mathbb{R}^3$. We discuss the generalization of the construction to semi-simple, possibly non simply connected, Lie groups based on their central extensions by suitable abelian Lie groups. Considering central extensions of Lie groups is the natural framework to deal with this problem, while the classification of the extensions needs to consider different types of cohomology including suitable group cohomologies. For some details on various relevant cohomologies in physics, see e.g \cite{brown,stor-wal,hoch-shap}.

{\bf{Notations}}: We will denote generically the involutions by the symbol $^\dag$. The actual nature of each involution should be clear from the context. $\mathcal{S}(\mathbb{R}^3)$ and $\mathcal{M}(\mathbb{R}^3)$ are respectively the algebra of Schwartz functions on $\mathbb{R}^3$ and its multiplier algebra. $\langle\cdot,\cdot\rangle$ is the hermitian product,i.e , $\langle f,g \rangle:=\int d^3x\ {\bar{f}}(x) g(x)$ for any $f,g\in\mathcal{M}(\mathbb{R}^3)$ (${\bar{f}}(x)$ is the complex conjugate of $f(x)$). $\tilde{f}(p)=\int d^3xf(x)e^{-ipx}$ is the Fourier transform of $f$. $\mathcal{L}(\mathcal{M}(\mathbb{R}^3))$ denotes the set of linear operators acting on $\mathcal{M}(\mathbb{R}^3)$.\\
We will deal with a family of deformations of $\mathbb{R}^3$, indexed by 3 functionals $f,\ g,\ l$ depending on the Laplacian of $\mathbb{R}^3$, $\Delta$, and a positive real parameter $\theta$ (the so-called deformation parameter), i.e $\mathbb{R}^3_{\theta,f,g,\ell}:=(\mathcal{F}(\mathbb{R}^3),\star_{\theta,f,g,\ell})$ ($\mathcal{F}(\mathbb{R}^3$ is a suitable linear space of functions to be characterized below) where $\star_{\theta,f,g,\ell})$ is the deformed product. To simplify the notations, any element of this family will be denoted by $\mathbb{R}^3_{\theta}(=(\mathbb{R}^3,\star))$, the actual nature of the objects indexing the family should be clear from the context.\\

\section{$\mathfrak{su}(2)$-noncommutativity and differential $^*$-representations.}\label{section2}
It is convenient to represent the abstract $^*$-algebra $\mathbb{A}[\hat{X}_\mu]$ generated by the self-adjoint operator coordinates $\hat{X}_\mu$ fulfilling $[\hat{X}_\mu,\hat{X}_\nu]=i2\theta\varepsilon_{\mu \nu}^{\hspace{11pt} \rho}\hat{X}_\rho$ ($\mu,\nu,\rho=1,2,3$) by making use of the poly-differential $^*$-representation
$\pi:\mathbb{A}[\hat{X}_\mu]\to\mathcal{L}(\mathcal{M}(\mathbb{R}^3))$, 
\begin{equation}
\pi:\hat{X}_\mu\mapsto\pi(\hat{X}_\mu)\dreq\hat{x}_\mu(x,\partial)=x_\nu\varphi^\nu_{\hspace{3pt}\mu}(\partial)+\chi_\mu(\partial), \label{defxhat}
\end{equation}
where the functionals $\varphi^\nu_{\hspace{3pt}\mu}(\partial)$ and $\chi_\mu(\partial)$ are viewed as formal expansions in the usual derivatives of $\mathbb{R}^3$, $\partial_\mu$, $\mu=1,2,3$. \\
Since by assumption $\pi$ is a morphism of $^*$-algebra, one has $[\hat{x}_\mu,\hat{x}_\nu]=i2\theta\varepsilon_{\mu \nu}^{\hspace{11pt} \rho}\hat{x}_\rho$ together with $\langle f,\hat{x}_\mu g\rangle=\langle \hat{x}_\mu f,g \rangle$ for any $f,g\in\mathcal{M}(\mathbb{R}^3)$, stemming from the self-adjointness of $\hat{x}_\mu$ so that $\hat{x}_\mu^\dag=\hat{x}_\mu$. By combining
these two latter conditions with \eqref{defxhat} and using $[x_\lambda , h(x,\partial)] = - \frac{\partial h}{\partial (\partial^\lambda)}$, which holds true for any functional $h$ of $x_\mu$ and $\partial^\mu$ together with $\partial^\dag_\mu=-\partial_\mu,\ h^\dag(\partial)={\bar{h}}(-\partial)$, a standard computation gives rise to the following functional differential equations constraining $\varphi^\nu_{\hspace{3pt}\mu}$ and $\chi_\mu$:
\begin{align}
i 2\theta \varphi_{\alpha \rho} &= \varepsilon_{\rho}^{\hspace{4pt} \mu \nu} \frac{\partial \varphi_{\alpha \mu}}{\partial (\partial_\beta)} \varphi_{\beta \nu},\label{master1}\\
\varphi^\dagger_{\alpha\rho} &= \varphi_{\alpha \rho} \label{master2}\\
i 2\theta \chi_\rho &= \varepsilon_{\rho}^{\hspace{4pt} \mu \nu} \frac{\partial \chi_\mu}{\partial (\partial_\alpha)} \varphi_{\alpha \nu}, \label{master3} \\
\frac{\partial \varphi^\dagger_{\alpha \rho}}{\partial (\partial_\alpha)} &= \chi_\rho - \chi_\rho^\dagger, \label{master4}
\end{align}
where use has been made of the algebraic relation $\delta_{\mu \gamma} \delta_\nu^{\hspace{4pt} \sigma} - \delta_\mu^{\hspace{4pt} \sigma} \delta_{\nu \gamma} = \varepsilon_{\mu \nu}^{\hspace{11pt} \rho} \varepsilon_{\rho \gamma}^{\hspace{11pt} \sigma}$.\\

In view of $\mathbb{R}^3_\theta\subsetneq U(\mathfrak{su}(2))\cong \mathbb{A}[\hat{X}_\mu]/[\hat{X}_\mu,\hat{X}_\nu]$, see e.g \cite{wal-16,jpw-16}, where $U(\mathfrak{su}(2))$ is the universal enveloping algebra of $\mathfrak{su}(2)$, there is a natural action of $SU(2)/\mathbb{Z}_2\simeq SO(3)$ on any $\mathbb{R}^3_\theta$. This selects $SO(3)$-equivariant $^*$-representations among those defined by \eqref{master1}-\eqref{master4}. A mere application of the Schur-Weyl decomposition theorem shows that the $SO(3)$-equivariance of the representation can be achieved whenever the functionals $\varphi^\mu_\nu$ and $\chi_\mu$ have the following form:
\begin{equation} 
\varphi_{\alpha \mu} (\partial) = f(\Delta) \delta_{\alpha \mu} + g(\Delta) \partial_\alpha \partial_\mu + i h(\Delta) \varepsilon_{\alpha\mu}^{\hspace{11pt} \rho} \partial_\rho,\label{polynomial_phi}
\end{equation}
\begin{equation} 
\chi_\mu(\partial) = \ell (\Delta) \partial_\mu \ \label{polynomial_chi},
\end{equation}
where the {\it{real}} $f(\Delta)$, $g(\Delta)$,  $h(\Delta)$ and {\it{complex}} $\ell(\Delta)$ $SO(3)$-invariant functionals are constrained by \eqref{master1}-\eqref{master4}.\\
By solving the constraints, one easily finds that the admissible solutions (i.e those $\hat{x}_\mu$ admitting an expansion of the form $\hat{x}_\mu=x_\mu+\mathcal{O}(\theta)$) are such that $h=\theta$ and form a family indexed by 3 functional $f$, $g$ and $\ell$ defined by 
\begin{equation} \label{general_rep}
\hat{x}_\mu = x^\alpha \left[ f(\Delta) \delta_{\alpha \mu} + g(\Delta) \partial_\alpha \partial_\mu + i\theta \varepsilon_{\alpha \mu}^{\hspace{11pt} \rho} \partial_\rho \right] + \ell(\Delta) \partial_\mu,
\end{equation}
\begin{align}
2\left[(f+g\Delta)' + g \right] &= \ell + \ell^\dagger \ , \label{1st_condition} \\
2(f+g\Delta)f' &= gf + \theta^2\label{2nd_condition}.
\end{align}
An interesting subfamily of poly-differential $^*$-representations arises whenever $f+g\Delta=1$, so that, setting $g(\Delta)\deq\frac{\theta^2}{3} G(2\theta^2 \Delta)$,  \eqref{1st_condition}, \eqref{2nd_condition} reduce to
\begin{eqnarray}
l+l^\dag&=&2g(\Delta) \ ,\label{kv-chi}\\
0&=&2t \frac{dG}{dt} + 3\left(G(t)+1 \right) - \frac{t}{6} G^2(t),\ \ \label{equadiff-reduced} ,
\end{eqnarray}
for which the Ricatti equation \eqref{equadiff-reduced} is solved by 
$G(t) = -6\sum_{n=1}^\infty \frac{2^n B_{2n}}{(2n)!} t^{n-1}$ where $B_n$ are Bernoulli numbers. The resulting subfamily{\footnote{Notice that consistency with \eqref{polar-dec} below will require $\ell$ to be real. }} takes the form
\begin{equation}
\hat{x}_\mu=x^\alpha\left[(1-g(\Delta)\Delta)\delta_{\alpha\mu}+g(\Delta)\partial_\alpha\partial_\mu+
i\theta\varepsilon_{\alpha\mu}^{\hspace{11pt}\rho}\partial_\rho
\right]+g(\Delta)\partial_\mu . \label{kv-star}
\end{equation}
We will derive from this subfamily a tracial star-product equivalent to the Kontsevich product for the Poisson manifold dual to the Lie algebra $\mathfrak{su}(2)$.\\

\section{Quantization maps, deformations of $\mathbb{R}^3$ and extensions.}
Let $Q$ denotes the quantization map, i.e an invertible $^*$-algebra morphism
\begin{equation}
Q: (\mathcal{M}(\mathbb{R}^3),\star) \to (\mathcal{L}(\mathcal{M}(\mathbb{R}^3)),\cdot)\label{quant-map},
\end{equation}
where "." is the product between differential operators omitted from now on, such that for any $f,\ g\in\mathcal{M}(\mathbb{R}^3)$,
\begin{equation}
f\star g \deq Q^{-1}\left(Q(f)Q(g)\right),\ Q(1)=\bbone,\ Q(\bar{f})=\left(Q(f)\right)^\dag,\label{alg-morph}
\end{equation}
\begin{equation}
Q(f)\rhd1=f(x) \ ,\label{Q-unitaction}
\end{equation}
so that $Q^{-1}\left(Q(f)\right)=Q(f) \rhd 1$, where "$\rhd$" is the left action of operators. Therefore the star-product can be expressed as
\begin{equation}
(f\star g)(x)=\int \frac{d^3p}{(2\pi)^3}\frac{d^3q}{(2\pi)^3}\tilde{f}(p)\tilde{g}(q)Q^{-1}\left(E_p(\hat{x})E_q(\hat{x}) \right) \ , \label{star-definition}
\end{equation}
for any $f,g\in\mathcal{M}(\mathbb{R}^3)$, where the quantized plane waves $E_p(\hat{x})$ are defined by
\begin{equation}
E_p(\hat{x}) \deq Q(e^{ipx}).\label{nc-planew}
\end{equation}
The quantized plane waves \eqref{nc-planew} define a map $E:SU(2)\to\mathcal{L}(\mathcal{M}(\mathbb{R}^3))$, 
\begin{equation}
E:g\mapsto E(g):=E_p(\hat{x}), \label{themap}
\end{equation}
\begin{equation}
E(g^\dag)=E^\dag(g) \ , \label{themapdag}
\end{equation}
for any $g\in SU(2)$. Using polar decomposition, one can write
\begin{equation}
E(g)=U(g)|E(g)| \ , \label{polar-dec}
\end{equation}
where $|E(g)|:=\sqrt{E^\dag(g)E(g)} \neq 0$ and the unitary operator $U:SU(2)\to\mathcal{L}(\mathcal{M}(\mathbb{R}^3))$ can be expressed as \begin{equation}
U(g)=e^{i\xi_g^\mu\hat{x}_\mu}, 
\end{equation}
in view of the Stone's theorem, where $\xi_g^\mu\in\mathbb{R}$. Then, the 
Baker-Campbell-Hausdorff formula for $\mathfrak{su}(2)$ gives 
\begin{equation}
e^{i\xi_{g_1}\hat{x}}e^{i\xi_{g_2}\hat{x}}=e^{iB(\xi_{g_1},\xi_{g_2})\hat{x}} \ ,\label{expans-bak}
\end{equation}
where the infinite expansion $B(\xi_{g_1},\xi_{g_2})$ fulfills 
\begin{equation}
B(\xi_{g_1},\xi_{g_2})=-B(-\xi_{g_2},-\xi_{g_1}) , \ B(\xi_g,0)=\xi_g \ .
\end{equation}
Observe that $U(g)$ and $E(g)$ define representations of $SU(2)$. Then, one has for any $g_1,g_2\in SU(2)$
\begin{equation}
U(g_1)U(g_2)=U(g_1g_2) \ , \label{projectif-su2}
\end{equation}
which holds true (up to unitary equivalence) as a mere application of the Wigner theorem to $SU(2)$, while we demand
\begin{equation}
E(g_1)E(g_2)=\Omega(g_1,g_2)E(g_1g_2)\label{projectif-caracter}
\end{equation}
where $\Omega(g_1,g_2)$ is to be determined. Combining $E(g^\dag g)=E(\bbone)=\bbone$ with Eqn. \eqref{projectif-caracter}, one obtains $E(g^\dag)E(g)=\Omega(g^\dag,g)\bbone$, for any $g\in SU(2)$. Therefore $\vert E(g)\vert=\sqrt{\Omega(g^\dag,g)}\bbone$, so that
\begin{equation}
\omega_g:=\sqrt{\Omega(g^\dag,g)} \in \mathbb{R},\ \omega_g>0,\label{positiv-omega}
\end{equation}
together with
\begin{equation}
[|E(g)|,U(g)]=0. \label{cond-central}
\end{equation}
Using \eqref{cond-central} and \eqref{polar-dec}, one get
\begin{equation}
E(g_1)E(g_2)=|E(g_1)||E(g_2)|U(g_1g_2)=|E(g_1)||E(g_2)||E(g_1g_2)|^{-1}E(g_1g_2) \ , \label{intermediaire}
\end{equation}
where the 2nd equality stems from \eqref{projectif-su2}, which combined with the expression for $|E(g)|$ yields
\begin{equation}
E(g_1)E(g_2)=(\omega_{g_1}\omega_{g_2}\omega^{-1}_{g_1g_2})E(g_1g_2) \ , \label{planew-multiplic}
\end{equation}
where
\begin{equation}
E(g_1g_2)=\omega_{g_1g_2}e^{iB(\xi_{g_1},\xi_{g_2})\hat{x}}.
\end{equation}
Note that \eqref{planew-multiplic} insures the associativity of the star-product \eqref{star-definition}, since the 2-cocycle $\Omega(g_1,g_2):=\omega_{g_1}\omega_{g_2}\omega^{-1}_{g_1g_2}$ obeys
$\Omega(g_1,g_2)\Omega(g_1g_2,g_3)=\Omega(g_1,g_2g_3) \Omega(g_2,g_3)$, for any $g_1,g_2,g_3\in SU(2)$. From eqn. \eqref{projectif-su2}, one infers that any unitary equivalent representations, $U$ and $U^\prime$ correspond to unitary equivalent products. Indeed, one has $U^\prime(g)=e^{i\gamma(g)}U(g)=e^{i\gamma(g)}e^{i\xi_g\hat{x}} $, where $\gamma$ is a real function, which implies the following equivalence relation $T(f\star^\prime g)=Tf\star Tg$ where $T$ is defined by $E^\prime_k(\hat{x})\equiv Q^\prime(e^{ikx}):=Q\circ T(e^{ikx})=e^{i\gamma(k)}Q(e^{ikx})=e^{i\gamma(k)}E_k(\hat{x}).$
Informally speaking, the star-product which will be determined in a while is essentially unique up to unitary equivalence. \\

This result can be understood \cite{tp-jcw17} within the more general framework of central extensions of Lie groups which in addition offers a convenient way for generalizations of the present work. In the following $\mathcal{A}$ is a 1-d abelian Lie group which we will assumed to be $\mathcal{A}=\mathbb{R}/\mathcal{D}$ where $\mathcal{D}$ is a discrete group of $\mathbb{R}$, i.e $\mathcal{D}=p\mathbb{Z}$. Let $G$ be a connected Lie group with Lie algebra $\mathfrak{g}$ whose action on $\mathcal{A}$, $\rho:G\times \mathcal{A}\to\mathcal{A}$, is assumed to be trivial ($\rho(g,a)=a$, for any $g\in G$, $a\in\mathcal{A}$). $\mathcal{E}$ is a central extension of $G$ if $\mathcal{A}$ is isomorphic to a subgroup of the center of $G$, $\mathcal{Z}(G)$ and $G\simeq \mathcal{E}/\mathcal{A}$ as group isomorphism. Recall that inequivalent central extensions of groups encoded by the short exact sequence $\bbone\to\mathcal{A}\to\mathcal{E}\overset{\pi}{\to} G\to \bbone$ (where $\pi$ is the canonical projection, with in addition $Im(\mathcal{A})\subset\mathcal{Z}(\mathcal{E})$), are classified by $H^2(G,\mathcal{A})$, the 2nd group of the cohomologie of $G$ with values in $\mathcal{A}$. Actually, up to additional technical requirements, $\mathcal{E}$ defines a principal fiber bundle over $G$ with 1-form connection and structure group $\mathcal{A}$. For {\it{simply connected}} Lie groups $G$, one has
\begin{equation}
H^2(G,\mathbb{R}/\mathcal{D})\simeq H^2_{alg}(\mathfrak{g},\mathbb{R}), 
\end{equation}
see e.g \cite{hoch-shap}, where $H^2_{alg}(\mathfrak{g},\mathbb{R})$ is the 2nd group of (real) cohomology of the Lie algebra $\mathfrak{g}$. \\
When $G$ is in addition semi simple, which is for instance the case for $SU(2)$, one has $H^2_{alg}(\mathfrak{g},\mathbb{R})=\{0\}$ so that $H^2(G,\mathbb{R}/\mathcal{D})$ is trivial. When $\mathcal{D}=\mathbb{Z}$, one has $\mathbb{R}/\mathcal{D}=U(1)$, hence the triviality of $H^2(SU(2),U(1))$ which explains the uniqueness of the above star-products up to unitary equivalence. The above conclusion extends to the central extension of any semi-simple and simply connected Lie group by $\mathbb{R}/\mathcal{D}$ so that the extension of the present construction to spaces with noncommutativity based on the corresponding Lie algebra should produce uniqueness of the star-product (up to equivalence).\\
When $G$ is semi-simple but not necessarily simply connected, its  inequivalent central extensions by $\mathbb{R}/\mathcal{D}$ are classified (up to additional technical requirements) by $H^1_{\hat{C}}(G,\mathbb{R}/\mathcal{D})$, where $H^\bullet_{\hat{C}}$ denotes the Cech cohomology. The following isomorphism holds true
\begin{equation}
H^1_{\hat{C}}(G,\mathbb{R}/\mathcal{D})\simeq Hom(\pi_1(G)\to\mathbb{R}/\mathcal{D}).
\end{equation}
Now assume $G=SL(2,\mathbb{R})$ and $\mathcal{D}=\mathbb{Z}$ so that $\mathbb{R}/\mathcal{D}\simeq U(1)$. The use of Iwazawa decomposition yields $SL(2,\mathbb{R})\simeq \mathbb{R}^2\times \mathbb{S}^1$, which combined with $\pi_1(X\times Y)=\pi_1(X)\times\pi_1(Y)$ for any topological spaces $X$ and $Y$ implies
$\pi_1(SL(2,\mathbb{R}))\simeq\mathbb{Z}$. Hence, $Hom(\mathbb{Z}\to U(1))\simeq U(1)$ which classifies the inequivalent extensions of $SL(2,\mathbb{R})$ by $U(1)$. Accordingly, one expect that the extension of the present construction of $SL(2,\mathbb{R})$ gives rise to inequivalent classes of star-products.\\

\section{Quantized plane waves for $\mathbb{R}^3_\theta$}

The determination of \eqref{themap} which takes the generic form
\begin{equation}
E_p(\hat{x})=\omega(p)e^{i\xi(p)\hat{x}} \label{generalform-ncexpo} ,
\end{equation}
can be achieved through a standard albeit tedious computation whose main steps are summarized now. \\
First, using $e^{i\xi(p)\hat{x}}\rhd 1 = \frac{e^{ipx}}{\omega(p)}$, one easily infers that
\begin{equation}
e^{-i\xi(\lambda p) \hat{x}} \partial_\mu e^{i\xi(\lambda p) \hat{x}} = (i\lambda p_\mu) \bbone,\label{megaplus1}
\end{equation}
which holds for any $\lambda\in\mathbb{R}$. Then, the combination of $[\partial_\mu, \hat{x}_\nu] = [\partial_\mu,x^a\varphi_{a\nu}] = \varphi_{\mu \nu}$ with the functional derivative of \eqref{megaplus1} with respect to $\lambda$ yields
\begin{equation}\label{diff-xi}
\varphi_{\mu\nu}(i\lambda p) \frac{d}{d\lambda} \left[ \xi^\nu(\lambda p) \right] = p_\mu.
\end{equation}
From $SO(3)$-covariance requirement, it can be shown \cite{tp-jcw17} that this latter functional differential equation is solved by
\begin{eqnarray} 
\xi^\mu(p) &=& \int_0^1 d\lambda (\varphi^{-1})^{\mu\nu}_{\vert_{i\lambda p}} p_\nu,\label{solution-xi}\\
(\varphi^{-1})^{\mu\nu}(ip) &=& \frac{1}{f^2+\theta^2 p^2} \left( f \delta^{\mu\nu} + 2f'p^\mu p^\nu + \theta \varepsilon^{\mu \nu \rho} p_\rho \right) \ , \label{phi-inverse}
\end{eqnarray}
where $f$ and $f^\prime$, \eqref{general_rep}-\eqref{2nd_condition}, depend on $(-p^2)$. $\xi_\mu$ can be verified to be an injective antisymmetric real-valued function. Finally, Eqn. \eqref{solution-xi} can be recast as a Volterra integral given by
\begin{equation} 
\xi^\mu(p) = \int_{-p^2}^0 \frac{dt}{2\vert p \vert\sqrt{-t}}\ \frac{f(t)-2tf^\prime(t)}{f^2(t)-\theta^2 t}p^\mu .\label{volterra-xi}
\end{equation} 
It can be verified that \eqref{volterra-xi} simplifies to $\xi^\mu=p^\mu$ when use it made of the subfamily of $^*$-representations \eqref{kv-star}. \\
To determine $\omega(p)$, one observes that
\begin{equation}
\frac{d}{d \lambda} \left[ e^{i\xi(\lambda p) \hat{x}} \right]= i (\varphi^{-1})^{\mu\nu}_{\vert_{i\lambda p}}  p_\nu \hat{x}_\mu e^{i\xi(\lambda p) \hat{x}} \ ,
\end{equation}
where \eqref{solution-xi} has been used, implying
\begin{equation}
\frac{d}{d \lambda} \left[ e^{i\xi(\lambda p) \hat{x}} \right] \rhd 1 = i \left( x^\nu + \chi_\mu (\varphi^{-1})^{\mu\nu}_{\vert_{i\lambda p}} \right) p_\nu \frac{e^{i\lambda px}}{\omega(\lambda p)} \ .
\end{equation}
Besides, the following relation holds true
\begin{equation}
\frac{d}{d \lambda} \left[ \frac{e^{i\lambda px}}{\omega(\lambda p)} \right] = \left( ix^\nu p_\nu - \frac{1}{\omega(\lambda p)} \frac{d}{d \lambda} \left[ \omega(\lambda p) \right] \right) \frac{e^{i\lambda px}}{\omega(\lambda p)}.
\end{equation}
Now, one can verify that $\frac{d}{d\lambda} \left[ \hat{A} f(x) \right] = \frac{d\hat{A}}{d\lambda} f(x)$ where $\hat{A}$ and $f$ are any operator and function suitably chosen for the latter expression to be well defined. Hence
\begin{eqnarray}
\frac{d}{d\lambda} \left[ e^{i\xi(\lambda p) \hat{x}} \rhd 1 \right] &=& \frac{d}{d\lambda} \left[ e^{i\xi(\lambda p) \hat{x}} \right] \rhd 1,\\
i \left( x^\nu + \chi_\mu (\varphi^{-1})^{\mu\nu}_{\vert_{i\lambda p}} \right) p_\nu &=& ix^\nu p_\nu - \frac{1}{\omega(\lambda p)} \frac{d}{d \lambda} \left[ \omega(\lambda p) \right] \ ,
\end{eqnarray}
implying
\begin{equation}
\frac{1}{\omega(\lambda p)} \frac{d}{d \lambda} \left[ \omega(\lambda p) \right] = - i \chi_\mu (\varphi^{-1})^{\mu\nu}_{\vert_{i\lambda p}} p_\nu \ ,
\end{equation}
which is solved by \cite{tp-jcw17} $\omega(p) = e^{-i \int_0^1 d\lambda \ \chi_\mu(i\lambda p) (\varphi^{-1})^{\mu\nu}_{\vert_{i\lambda p}} p_\nu}$. This latter expression can be recast as a Volterra integral given by 
\begin{equation} \label{volterra-omega}
\omega(p) = e^{\int_{-p^2}^0 dt\ \frac{f(t)-2tf^\prime(t)}{f^2(t)-\theta^2 t}\ell(t)}.
\end{equation}
Consistency with \eqref{positiv-omega} requires $\omega(p)$ to be a positive real quantity therefore constraining $\ell$ to be a real functional, $\ell^\dag=\ell$. It follows that \eqref{1st_condition} reduces to $\ell=(f+g\Delta)^\prime+g$, which thus constraints $\ell$ once $f$ and $g$ fulfilling \eqref{2nd_condition} are obtained.\\

To summarize the above analysis, given a $^*$-representation belonging to the family \eqref{general_rep}-\eqref{2nd_condition}, eqns. \eqref{volterra-xi} and \eqref{volterra-omega} fully characterize the corresponding quantization map $Q$ \eqref{quant-map}, \eqref{Q-unitaction} together with its star-product \eqref{star-definition}. Two remarks are in order:\\
i) One can show \cite{tp-jcw17} that for the family of poly-differential $^*$-representations considered in this note, $Q$ {\it{cannot}} be the Weyl quantization map $W$related to the symmetric ordering for operators. In fact, the only poly-differential representation compatible with the Weyl quantization is defined by $\hat{x}_\mu=x^\alpha\left((1-g\Delta)\delta_{\alpha\mu}+g\partial_\alpha\partial_\mu+i\theta
\varepsilon_{\alpha\mu\rho}\partial_\rho\right) $ (in particular $\chi=0$), where $g$ has been defined above, which however is not a $^*$-representation.\\
ii) One can verify that the star-product used in \cite{lagraa} and some ensuing works does not belong to the general family of star-products related to \eqref{general_rep}. The former star-product, defining a particular deformation of $\mathbb{R}^3$ called $\mathbb{R}^3_\lambda$, can be related to the Wick-Voros product \cite{wick-voros, galuccio, fedele} stemming from a twist. So far, whether or not the present family of star-products also admits a representation in terms of a twist is not known.

\section{Tracial star-product and related quantum field theories}\label{subsection34}

In this section, we restrict ourselves to the particular subfamily of $^*$-representations \eqref{kv-star}. By combining this latter with eqns. \eqref{volterra-xi} and \eqref{volterra-omega}, one easily finds that the quantized plane waves are defined by $Q(e^{ipx})\equiv E_p(\hat{x}) = \left( \frac{\sin(\theta |p|)}{\theta |p|} \right)^2 e^{ip\hat{x}}$, from which follows
\begin{equation}
(f\star_Q g)(x) = \int \frac{d^3p}{(2\pi)^3}\frac{d^3q}{(2\pi)^3}\tilde{f}(p)\tilde{g}(q) \mathcal{W}^2(p,q) e^{iB(p,q)x} \ ,
\end{equation}
for any $f,g \in \mathcal{M}(\mathbb{R}^3)$, with $\mathcal{W}(p,q) \deq \frac{|B(p,q)|}{\theta |p||q|}\frac{\sin(\theta |p|)\sin(\theta |q|)}{\sin(\theta |B(p,q)|)}$ where $B(p,q)$ is given by \eqref{expans-bak}. Now, introduce a new quantization map $\mathcal{K}:\mathcal{M}(\mathbb{R}^3) \to \mathcal{L}(\mathcal{M}(\mathbb{R}^3))$ through
\begin{equation}
\mathcal{K} \deq Q \circ H,
\end{equation}
where the operator $H$ acting on $\mathcal{M}(\mathbb{R}^3)$ is given by
\begin{equation} \label{Kontsevich}
H \deq \frac{\theta \sqrt{\Delta}}{\sinh(\theta \sqrt{\Delta})},
\end{equation}
and such that
\begin{equation}
H(f\star_\mathcal{K}g)=H(f)\star_QH(g),
\end{equation}
for any $f,g \in \mathcal{M}(\mathbb{R}^3)$. One obtains from a standard computation
\begin{equation}
\mathcal{K}(e^{ipx}) = \frac{\sin(\theta |p|)}{\theta |p|} e^{ip\hat{x}} \label{checkpoint2},
\end{equation}
from which it is easy to obtain finally
\begin{equation}
(f\star_\mathcal{K}g)(x)=\int \frac{d^3p}{(2\pi)^3}\frac{d^3q}{(2\pi)^3}\tilde{f}(p)\tilde{g}(q) \mathcal{W}(p,q) e^{iB(p,q)x} \ , \label{kontsev-product}
\end{equation}
for any $f,g \in \mathcal{M}(\mathbb{R}^3)$, where $\mathcal{W}(p,q)$ has been given above. \\

The star-product $\star_\mathcal{K}$ \eqref {kontsev-product} is nothing but the Kontsevich product \cite{deform} related to the Poisson manifold dual to the finite dimensional Lie algebra $\mathfrak{su}(2)$,. This can be realized \cite{tp-jcw17} by noticing that \eqref{checkpoint2} can be recast into the form $\mathcal{K}(e^{ipx})=W(j^{\frac{1}{2}}(\Delta)(e^{ipx}))$ where $W$ is the Weyl quantization map, $W(e^{ipx})=e^{ip\hat{x}}$ and
\begin{equation}
j^{\frac{1}{2}}(\Delta)=\frac{\sinh(\theta \sqrt{\Delta})}{\theta \sqrt{\Delta}} \ , \label{duflo}
\end{equation}
is the Harish-Chandra map \cite{harish}, \cite{duflo}. Hence,
\begin{equation}
\mathcal{K}=W\circ j^{\frac{1}{2}}(\Delta) \ , \label{deriv-konts}
\end{equation}
which coincides with the Kontsevich product in the present case, see e.g \cite{q-grav}. Note that \eqref{Kontsevich} and \eqref{duflo} imply
\begin{equation}
j^{\frac{1}{2}}(\Delta)=H^{-1},
\end{equation}
so that $H$ \eqref{Kontsevich} is the inverse of the Harish-Chandra map. The star-product $\star_\mathcal{K}$ \eqref {kontsev-product} is tracial with respect to the trace functional which in the present case is simply defined by the Lebesgue integral on $\mathbb{R}^3$. Namely, one has
\begin{equation}
\int d^3 x (f \star_\mathcal{K} g)(x) = \int d^3 x f(x) g(x).
\end{equation}

This last interesting property can be exploited to built NCFT admitting standard (i.e commutative) massive real or complex scalar field theories with quartic interaction as formal commutative limits \cite{jpw-16}, \cite{KV-15}, namely
\begin{eqnarray}
S_1&=&\int d^3x\big[\frac{1}{2}\partial_\mu\phi\star_\mathcal{K}\partial_\mu\phi+\frac{1}{2}m^2\phi\star_\mathcal{K}\phi+\frac{\lambda}{4!}\phi\star_\mathcal{K}\phi\star_\mathcal{K}\phi\star_\mathcal{K}\phi\big]\label{real-clasaction},\\
S_2&=&\int d^3x\big[\partial_\mu\Phi^\dag\star_\mathcal{K}\partial_\mu\Phi+m^2\Phi^\dag\star_\mathcal{K}\Phi+
{\lambda}\Phi^\dag\star_\mathcal{K}\Phi\star_\mathcal{K}\Phi^\dag\star_\mathcal{K}\Phi\big]\label{complx-clasaction}.
\end{eqnarray}
It turns out that \eqref{real-clasaction} and \eqref{complx-clasaction} do not have (perturbative) UV/IR mixing, as shown in \cite{jpw-16}. Indeed, the relevant contributions to the 2-point function can be written as $\Gamma_2^{(I)}=\int d^3x\ \phi(x)\phi(x)\omega_I$ and $\Gamma_2^{(II)}=\int \frac{d^3k_1}{(2\pi)^3}\frac{d^3k_1}{(2\pi)^3}\ \tilde{\phi}(k_1)\tilde{\phi}(k_2)\omega_{II}(k_1,k_2)$, with
\begin{eqnarray}
\omega_{I}&\sim&\frac{4}{\theta^2}\int\frac{d^3p}{(2\pi)^3}\ \frac{\sin^2(\frac{\theta}{2}|p|)}{p^2(p^2+m^2)}=\frac{1-e^{-\theta m}}{2m\pi\theta^2}\\
\omega_{II}&\sim&\int d^3x\frac{d^3p}{(2\pi)^3}\ \frac{1}{p^2+m^2}(e^{ipx}\star_{\mathcal{K}}e^{ik_1x}\star_{\mathcal{K}}e^{-ipx}\star_{\mathcal{K}}e^{ik_2x}).
\end{eqnarray}
When $\theta\ne0$, $\omega_I$ is obviously finite even for $m=0$ while  $\omega_{II}(0,k_2)\sim\delta(k_2)\omega_I$. Similar expression holds for $\omega_{II}(k_1,0)$ so that no IR singularity occurs within \eqref{real-clasaction}. This signals the absence of UV/IR mixing. One can check \cite{jpw-16} the UV one-loop finiteness of $\omega_{II}$. Note that similar conclusions hold true for the complex scalar field case \eqref{complx-clasaction}. \\

The origin of the absence of UV/IR mixing as well as the present mild (finite) UV behaviour (which should extend to all orders) is likely due to the Peter-Weyl decomposition{\footnote{This should hold provided the kinetic operator has a reasonable behaviour, i.e has compact resolvant insuring a sufficient decay of the propagator. }}of the algebra modeling the noncommutative space, i.e $\mathbb{R}^3_\theta=\oplus_{j\in\frac{\mathbb{N}}{2}}\mathbb{M}_{2j+1}(\mathbb{C})$, which reflects the relationship between the algebra and the convolution algebra of $SU(2)$, see \cite{wal-16}. This is particularly apparent in a class of NCFT \cite{vitwal}, with however kinetic operators different from the usual Laplacian used here, for which the theory splits into an infinite tower of (matrix) field theories, each on a finite geometry with the radius $\sim j$ of $\mathbb{M}_{2j+1}(\mathbb{C})$ serving as a natural cut-off. The use of the standard Laplacian in 
\eqref{real-clasaction}, \eqref{complx-clasaction} complicates the analysis of the UV behaviour to arbitrary orders but one can reasonably conjecture that both \eqref{real-clasaction} and \eqref{complx-clasaction} NCFT are (UV) finite to all orders.\\

An interesting issue would be to extend the above analysis to the case of noncommutative gauge theories whose commutative limit would reproduce the usual gauge (Yang-Mills) theory on $\mathbb{R}^3$. Such an extension would presumably exclude the choice of a derivation-based differential calculus \cite{mdv-jcw} since this latter would produce natural Laplacians without (analog of) radial dependence. Note that one proposal aiming to include radial dependence presented in \cite{fedele} amounts to enlarge the initial algebra by incorporating the deformation parameter itself. \\

\vskip 4 true cm

{\bf{Acknowledgments:}} J.-C. Wallet is grateful to F. Besnard, N. Franco and F. Latr\'emoli\`ere for discussions on various topics related to the present work. T. Poulain thanks COST Action MP1405 QSPACE for partial financial support. This work is partially supported by H2020 Twinning project No. 692194, “RBI-T-WINNING”.

\end{document}